\documentclass[journal]{IEEEtran}

\pdfoutput=1

\usepackage{graphicx}
\usepackage{amsmath}
\ifCLASSINFOpdf
 \else
 
\fi

\begin{document}

\title{Social Engineering Resistant 2FA}

\author{Markus Jakobsson\\
{\tt markus@zapfraud-inc.com}}

\maketitle

\begin{abstract}
Attackers increasingly, and with high success rates, use social engineering techniques to circumvent second factor authentication (2FA) technologies, compromise user accounts and sidestep fraud detection technologies. We introduce a social engineering resistant approach that we term {\em device-aware} 2FA\footnote{Patents pending.}, to replace the use of traditional security codes. \end{abstract}

\IEEEpeerreviewmaketitle

\section{Introduction}

\IEEEPARstart{S}{ocial Engineering} has emerged as one of the principal threats against Internet security, 
and, by extension, against society~\cite{Krebs2016}.
Social engineering  corresponds to a spectrum of related abuses, all of which have 
one thing in common: The use of deception to sidestep existing security measures. 
Social engineering is at the heart of phishing attacks, for example, whether the attacker aims to siphon 
off funds or emails from
compromised accounts; it is also the principle underlying the operation of Business Email Compromise attacks, where criminals pose as trusted colleagues and request for payments to be made or data to be shared. In both of these attacks, 
one part of the social engineering involves posing as a trusted party. 

\begin{figure}[h!]
\centering
\includegraphics[width=2in]{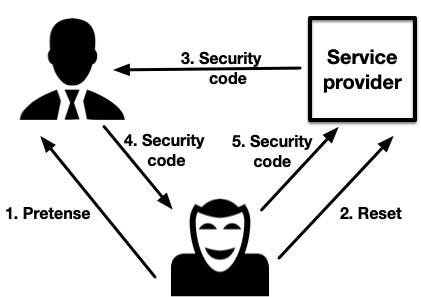}
\caption{The figure illustrates the typical flow of a social engineering attack on 2FA, in which an attacker uses
deception to trick a victim to forward a security code that enables the attacker to compromise the victim's account.
A related attack can be used against token-based 2FA solutions.
\label{Attackflow}}
\end{figure}

The security community's initial reaction to new social engineering attacks appears always to be awareness campaigns. 
While these are typically not very effective, they are quick to deploy and help contain the damage as technical countermeasures are developed and rolled out. Once deployed, the technical countermeasures reduce the profits of the criminals, forcing them to search for alternative ways of deceiving people. 
As a prime example of this, we find that 
the recent successes  of security technologies that identify and block identity deception have resulted in 
criminals instead 
 trying to compromise legitimate email  accounts. 
 The compromised accounts, in turn, are used to harvest information and to deceive other users -- namely contacts of the owners of the 
 compromised accounts. Currently deployed security systems, many of which focus on identifying 
 when an attacker masquerades as another user, do not flag emails from compromised accounts -- after all, these are emails from 
{\em legitimate} accounts, sent to people who in the past have communicated with these users. 
 Typically, the criminals compromise these ``launchpad'' accounts by stealing the credentials from the account owners. 
 One increasingly common approach of doing that is for the criminals to initiate a password reset for an 
 intended launchpad account, and then use 
  social engineering methods to trick the account owner to 
  send them the 
  password reset code.
  This attack, which is illustrated in 
  Fig.~\ref{Attackflow}, is 
  also commonly used to complete high-value transactions for accounts that have already been compromised, or to circumvent any other use of 2FA. 
 
\begin{figure}[h!]
\centering
\includegraphics[width=2.3in]{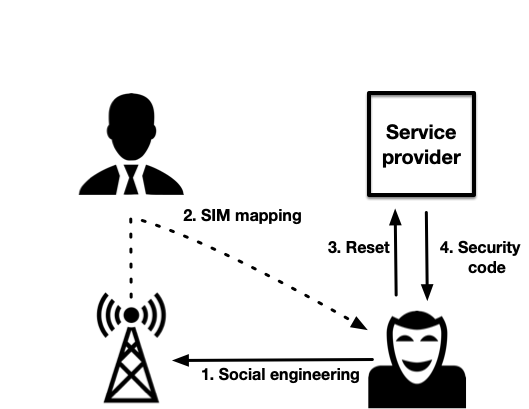}
\caption{The figure illustrates SIM-jacking, followed by the automated theft of a 2FA code. The attacker causes the carrier to map the victim's phone number to the attacker's SIM. After requesting a password reset from a service provider, the 2FA code is sent directly to the
attacker's phone.
\label{SIMjack}}
\end{figure}

Another common approach to compromise accounts  is {\em SIM-jacking} \cite{SIM, SIM24}. 
 In this attack, the criminal does not interact with the intended victim. Instead, he causes the victim's {\em carrier} to change the mapping between the victim's phone number and the associated SIM card  so that
 any call or SMS intended for the victim user instead get routed to the attacker.  This  is illustrated in 
  Fig.~\ref{SIMjack}.  
Once this is done, the criminal can gain access to any of the victim's accounts secured by SMS-based 2FA, simply by 
initiating password resets on behalf of the victim. As the service provider sends password reset codes to the victim's phone number, 
these are delivered to the criminal. This attack recently caused Twitter to make SMS-based 2FA {\em optional}  for their users ~\cite{twitter} --- however, it is unclear that the alternatives offered provide much better protection. 
 
 \vspace{-0em}
  
  In this paper, we introduce  a new type of SMS-based 2FA 
method, protecting both against attacks where the attacker uses social engineering to deceive his
victim and attacks based on SIM-jacking. 
 
\section{A Brief History of Social Engineering}

Social Engineering first became the material of mainstream media coverage in the mid-2000s, with the rapid rise 
of phishing attacks. In early phishing attacks, the criminals spoofed financial institutions by abusing weaknesses in the 
email protocol,  thereby making the phishing emails appear to have been sent by the impersonated financial 
institutions. The spoofed emails instructed
 would-be victims to 
click on hyperlinks and ``log in.'' 
At first, the response from the security community was that of awareness campaigns, commonly suggesting that the users look out for 
poorly spelled or poorly formatted emails. As this approach proved insufficient, it was soon 
 followed by efforts to quickly identify  phishing websites
referred to in the phishing emails, and to mount take-down efforts of these. A lack of procedure, however, initially 
resulted in typical take-down times on the order of a month, later coming down to hours or days \cite{mc07}. 
Given that typical user reactions commonly take place within minutes or hours \cite{socialphishing}, take-down was not 
a sustainable response, and, like awareness campaigns, failed to curb the abuse. 
The volume of phishing attacks only started to fall  in response to the rollout of DMARC \cite{DMARC}, an Internet standard addressing  email spoofing.

\begin{figure}[h!]
\centering
\includegraphics[width=2.9in]{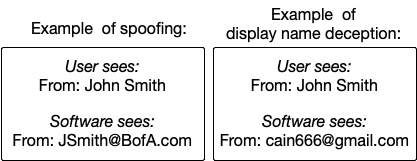}
\caption{The figure illustrates the difference between spoofing and the use of deceptive display names. In a spoofed email, the 
impersonated user appears as the apparent sender, but in an attack using deceptive display names, that is not so.  DMARC detects spoofed emails, but not emails with deceptive display names.}
\label{Spoof_DN}
\end{figure}

Starting around 2015, Business Email Compromise (BEC) attacks saw a similarly meteoric rise 
\cite{FBI18, FBI19} 
as phishing attacks had ten years earlier. In BEC attacks, criminals pose as trusted parties by using deceptive display names,  thereby  effectively
dodging the protection offered by DMARC by avoiding the use of email spoofing. (The difference between spoofing and deceptive display names is
illustrated in Figure~\ref{Spoof_DN}.)

\begin{figure}[h!]
\centering
\includegraphics[width=3.4in]{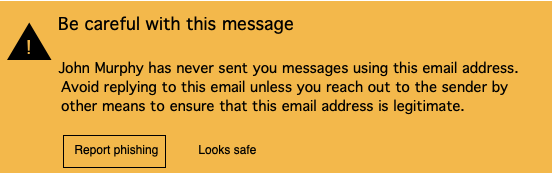}
\caption{The figure shows a Google-issued warning shown to a user receiving an email from a 
``new'' sender with the same display name as a (to the recipient) known sender. By drawing the attention of the recipient to the potential risk, abuses such as Business Email Compromise are countered.}
\label{googleBEC}
\end{figure}

At first, the response of the security community was, again, that of awareness campaigns. 
While  these still remain a commonly advocated line of defense, 
many email security companies have also introduced automated detection methods that identify high-risk emails
by determining when emails from non-trusted parties use display names of trusted parties and alert the recipient -- see, e.g.,  Figure~\ref{googleBEC}.
Like DMARC 
automated the detection of typical phishing attacks, the detection of deceptive display names 
has addressed BEC and related abuses. 
However, the success of these countermeasures is also causing the attacks to evolve, leading to the {\em next} wave of attacks.

 While criminals have shown no sign of abandoning social engineering, they have -- time after time -- demonstrated that they will 
  move on to new types of attacks once the 
security industry has deployed countermeasures that sufficiently hamper their efforts. Thus, the successes within 
the security industry, ironically, are what shape the criminal trends. 
The two most recent trends in social engineering both relate to account compromise, and while both attacks
are well understood, the deployed defenses have not yet matured beyond awareness campaigns. 

{\bf Launchpad Attacks.}
In one of these two types of attacks, 
termed a {\em launchpad attack} \cite{Jakobsson2019TheRT}, an attacker that has compromised an email account (of a ``launchpad user'') uses the compromised account 
to send out messages to contacts of the email account owner. These users, who are the {\em real} victims of the attack,
may be asked to make payments, share documents, or install software. Since the attacker uses a 
legitimate account of a trusted contact, 
both the victim and his security technologies
 decide (with a high likelihood) that the email is legitimate, making this attack very dangerous.

{\bf Account Compromise.} 
In the {\em other} social engineering attack currently on the rise (see, e.g., \cite{phonephish},) an attacker {uses social engineering for account compromise}, 
as opposed to using {compromised accounts for social engineering}. 
The two common approaches, as described in the introduction, is to attempt to trick the account owner to 
forward password reset codes, or to use SIM-jacking to steal the delivery channel for the password reset codes. 

The victim-facing attack has been partially addressed in a recent publication by Siadati et al. \cite{SIADATI201714}, while SIM-jacking has 
remained  unaddressed \cite{EmpStudy}, becoming 
a thorny security issue to many organizations \cite{SoHey}. 
Many launchpad attacks start out as account compromise attacks. However, account compromise attacks are used in many other 
types of abuses, such as in phishing attacks used to infiltrate organizations \cite{bypass}. 

This article focuses on improving the security of 2FA by introducing a new SMS-based 2FA paradigm, thereby 
protecting users against account compromise, and the attacks that stem from it.

\section{Solution Overview}
\label{sol}

The user experience
of our proposed solution closely resembles that of traditional methods. However, under the hood, our proposed 
method provides additional functionality: Using device identification methods, the 
 service provider (e.g., the website of the bank, social media provider or email service provider)
 determines whether a 2FA challenge is being forwarded or whether it is responded to by a recognized device associated with the 
account  of the 2FA challenge. 

\subsection{Device-Aware 2FA}

\begin{figure}[h!]
\centering
\includegraphics[width=3.2in]{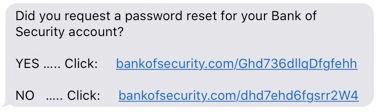}
\caption{The figure shown an example 2FA challenge with two clickable links. If the user clicks on either, the service provider receives a request from the user's browser, along with HTML cookies and other trackers, which are used to determine whether the click originated from a device recognized as belonging to the account owner.
While the SMS standard does not support HTML, most SMS apps will interpret SMS text and identify URLs.
\label{2FAexample}}
\end{figure}

Instead of sending traditional 2FA messages containing security codes, 
the service provider sends a 2FA challenge with one or more clickable links. 
An example 2FA challenge is shown in Fig.~\ref{2FAexample}.

\begin{figure}[h!]
\centering
\includegraphics[width=2.8in]{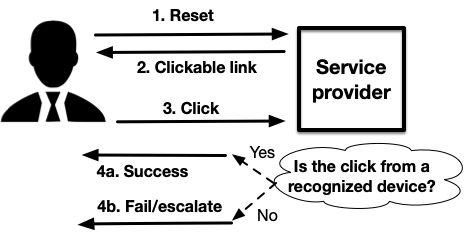}
\caption{The figure illustrates the flow for 2FA challenges. 
Instead of manually copying codes, the user only has to 
 {click}.
Escalation can be performed by 
 sending {\em another} 2FA challenge, e.g.,  to another registered phone number or email address associated with  the user, or to perform knowledge-based authentication.
\label{2FAflow2}}
\end{figure}

When the user clicks on a link, the service provider receives a webpage request from the user's browser, along with any HTML cookies (and other trackers) associated with the user device and the service provider domain. The service provider verifies that the click came from a
device that is recognized as belonging to the account owner. This makes the second factor authentication {\em device-aware}.
A device-aware 2FA process is illustrated in Figure~\ref{2FAflow2}.

\begin{figure}[h!]
\centering
\includegraphics[width=2.8in]{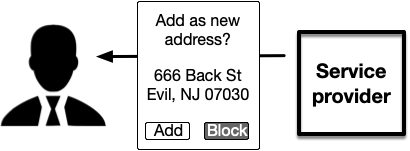}
\caption{The figure illustrates the action taken in response to a successful 2FA challenge.  The user is asked to review and approve the 
sensitive action that triggered the 2FA. The approval is taking place on the  device used to respond to the 2FA challenge, as opposed to on the device that initiated the
request for the sensitive action. 
\label{Success}}
\end{figure}

The 2FA is triggered by a request to perform a sensitive action -- 
such as changing a password or a delivery address, performing a large transfer of funds, or performing an action that is anomalous, based on the user's history. The 2FA challenge is said to {\em succeed} if the user's device is recognized. The user is then allowed to complete the sensitive action on
the same device that was used to respond to the challenge -- this ascertains that the user knows what action is being completed. An example is shown in Figure~\ref{Success}. 
If, on the other hand, the user's device is {\em not} recognized, then an escalation process is performed, if available; this may involve using {\em another} registered device to pass a 
second 2FA challenge or to perform knowledge-based authentication. Most organizations have escalation processes in place already, since 
the standard SMS-based 2FA also can fail. 

\subsection{Understanding Security}

With device-aware 2FA, traditional social engineering that aims to trick the intended victim to forward 2FA challenges to the attacker will no longer be meaningful. This is because the attacker's device will not be recognized as belonging to the account owner: The attacker does not have the cookies (and other identifiers) that the user has. SIM-jacking will also not succeed, for the same reasons. 

 Consider now an attacker who 
 performs an action that causes a 2FA challenge to be sent to an intended victim, and who then
 tricks the intended victim to click on the 2FA challenge. The service provider will consider the 2FA challenge successful, since a recognized device was used to respond to it; however, the account owner still has to approve or complete the sensitive action in order for this to be completed.
 For example, the account owner may have to approve a new address, as illustrated in  Figure~\ref{Success}, or may be requested to set a new 
  password.  
This completion process is performed on the device used to pass the 2FA challenge (e.g., the user's phone), as opposed to on the device that was used to initiate the 
request that caused the 2FA (e.g., the attacker's computer).  Therefore, the user will be aware of the action that is being performed.

\subsection{Recognizing Devices}
Device-aware 2FA takes advantage of device-identifying technologies that are already widely deployed, but uses them differently. These device-identifying technologies include various types of cookies placed by a website onto a device, such as HTML cookies, flash cookies and other types of cookies. These identifiers are typically not possible to steal by an attacker who has not already compromised the user's device -- which is 
more involved than to compromise 2FA  or separate accounts accessed from the device. 
 Moreover, almost all websites already use such device identifiers, whether for personalization or fraud detection. In fact, 2FA is often activated if a user attempts to log in from a device that is not recognized. 

Device-identifying technologies also include  
read-only characteristics such as  browser type and version installed, touch-screen support, system fonts installed, languages installed, screen size, color depth, time zone, and browser plug-in details. 
Service providers can also use additional 
 characteristics such as the user's network name, carrier name and information related to her geolocation.
While the digital fingerprints made up by these read-only identifiers aren't unique to each device, there are so many permutations of user hardware and software attributes that it is unlikely for two devices to share a common fingerprint. Also, while a competent attacker can be expected to steal {\em some} of the read-only identifiers by tricking his victim to click on a link to a website the attacker controls, it is unlikely for a typical attacker to successfully steal {\em all} of these identifiers. 

By using risk-based authentication methods, service providers can determine whether a given requested sensitive action should be allowed, given 
the available evidence that a device corresponds to a recognized device. 

\section{Special Case: New Device}
\label{new}

When the system receives a response to a 2FA challenge from a ``new'' device, it has to determine whether this
corresponds to an attack or to a user having replaced her phone (or other device) since the last 
time she used to to access the service provider. 
This problem
can be addressed using a variety of well-known escalation methods.

Here, escalation may correspond to asking the user to use another registered device to respond to a new
challenge; to respond to knowledge-based authentication (KBA) questions; or otherwise prove access to 
a resource (e.g., being able to determine the message associated with a 1-cent payment to a checking account known
to be associated with the user.) The exact escalation process depends on the type of service associated with the 
2FA challenge.

\begin{figure}[h!]
\centering
\includegraphics[width=3.4in]{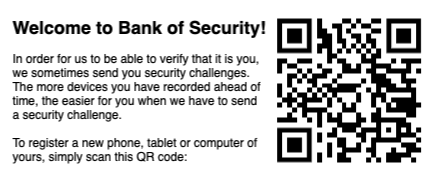}
\caption{The figure illustrates a possible device registration approach. As the user scans the QR code with her new device, a request is sent to a webpage whose URL is encoded using the QR code. This URL is unique to the user and the time at which the QR code is presented. As the request is received, the requesting device is profiled by the service provider.
\label{register}}
\end{figure}

An important system functionality is therefore to register new devices associated with a user account. 
Device profiles can be bootstrapped as users register devices during account setup, and by asking an 
already recognized (and logged-in) user to enroll new devices, as illustrated 
in Figure~\ref{register}. 

Devices can also be 
automatically registered. 
For example, if a user fails a 2FA challenge at first, due
to her device not being registered, and then {\em succeeds} during the escalation process, then the user can
be asked to confirm that she performed the failed access; if she did, the device associated with the 
failed access will be automatically registered. It is desirable for users to 
 have multiple devices registered; e.g., the user's computer, tablet and phone, and potentially also a device belonging to a trusted friend (where a reset from such a device may require that the user also answers knowledge-based questions.) 

\section{Discussion}
\label{ana}

We have described a technical solution that addresses a range of  social engineering attacks related to
2FA. Our solution is independent of the technology used for the delivery of the attack -- it works, for example, no matter 
whether the mode of operation of a criminal is to attempt to circumvent 2FA solutions in a manual manner (placing a phone call to the victim) or in a highly automated manner (sending ``phishing'' emails requesting the collaboration of the victim). It also addresses attacks such as SIM-jacking attacks in which the {\em delivery channel} is stolen.

Our approach is based on user devices being recognized by service providers -- a well understood problem that the industry has already solved in a very robust manner for reasons unrelated to 
2FA -- and offers a user experience that is sufficiently similar to existing methods for us to expect 
a simple adoption.

We believe that addressing today's vulnerabilities associated with 2FA technologies 
can have large positive benefits by eliminating a common vector of compromise. However, we are mindful of the
fact that social engineering is a moving target, and that protecting against one form of abuse entices 
the criminals to develop another. 

 Thus, we believe that  the security community must 
 remain vigilant and constantly on the lookout for new types of abuses. Once identified, it is important to 
develop {\em technical} solutions -- as opposed to solutions based on awareness -- to address the new-found abuses. 
 It is our belief that, while a useful second line of defense, user awareness
should not be the primary defense against social engineering.

\section*{Acknowledgment}
Thanks to Toan Nguyen for helpful feedback on an earlier version, and to Paul Sherer for help with the presentation.

\ifCLASSOPTIONcaptionsoff
  \newpage
\fi

\bibliographystyle{IEEEtran}
\bibliography{Device-Aware-2FA-Markus-Jakobsson}

\begin{thebibliography}{10}
\providecommand{\url}[1]{#1}
\csname url@samestyle\endcsname
\providecommand{\newblock}{\relax}
\providecommand{\bibinfo}[2]{#2}
\providecommand{\BIBentrySTDinterwordspacing}{\spaceskip=0pt\relax}
\providecommand{\BIBentryALTinterwordstretchfactor}{4}
\providecommand{\BIBentryALTinterwordspacing}{\spaceskip=\fontdimen2\font plus
\BIBentryALTinterwordstretchfactor\fontdimen3\font minus
  \fontdimen4\font\relax}
\providecommand{\BIBforeignlanguage}[2]{{%
\expandafter\ifx\csname l@#1\endcsname\relax
\typeout{** WARNING: IEEEtran.bst: No hyphenation pattern has been}%
\typeout{** loaded for the language `#1'. Using the pattern for}%
\typeout{** the default language instead.}%
\else
\language=\csname l@#1\endcsname
\fi
#2}}
\providecommand{\BIBdecl}{\relax}
\BIBdecl

\bibitem{Krebs2016}
B.~Krebs, ``{Russian 'Dukes' of Hackers Pounce on Trump Win},''
  \url{https://krebsonsecurity.com/2016/11/russian-dukes-of-hackers-pounce-on-trump-win/\#more-36921},
  November 16, 2016.

\bibitem{SIM}
D.~H. Kass, ``{SIM Card Swap Attacks: Will Lawsuit Pressure Two Factor
  Authentication?}''
  \url{https://www.msspalert.com/cybersecurity-services-and-products/mobile/sim-card-swap-attacks/},
  Aug 8, 2019.

\bibitem{SIM24}
R.~McMillan, ``{He Thought His Phone Was Secure; Then He Lost \$24 Million to
  Hackers},''
  \url{https://www.wsj.com/articles/he-thought-his-phone-was-secure-then-he-lost-24-million-to-hackers-11573221600},
  Nov. 8, 2019.

\bibitem{twitter}
G.~Kumparak, ``Twitter will finally let you turn on two-factor authentication
  without giving it a phone number,''
  \url{https://techcrunch.com/2019/11/21/twitter-will-finally-let-you-turn-on-two-factor-authentication-without-giving-it-a-phone-number/},
  Nov 21, 2019.

\bibitem{mc07}
\BIBentryALTinterwordspacing
T.~Moore and R.~Clayton, ``Examining the impact of website take-down on
  phishing,'' in \emph{Proceedings of the Anti-Phishing Working Groups 2nd
  Annual ECrime Researchers Summit}, ser. eCrime ?07.\hskip 1em plus 0.5em
  minus 0.4em\relax New York, NY, USA: Association for Computing Machinery,
  2007, p. 1?13. [Online]. Available:
  \url{https://doi.org/10.1145/1299015.1299016}
\BIBentrySTDinterwordspacing

\bibitem{socialphishing}
T.~N. Jagatic, N.~A. Johnson, M.~Jakobsson, and F.~Menczer, ``{Social
  Phishing},'' \emph{Commun. {ACM}}, vol.~50, no.~10, pp. 94--100, 2007.

\bibitem{DMARC}
M.~Moorehead, ``{How to Explain DMARC in Plain English},''
  \url{https://blog.returnpath.com/how-to-explain-dmarc-in-plain-english/},
  July 20, 2015.

\bibitem{FBI18}
{Federal Bureau of Investigation}, ``{Business E-mail Compromise: The 12
  Billion Dollar Scam},'' \url{www.ic3.gov/media/2018/180712.aspx}, July 12,
  2018.

\bibitem{FBI19}
------, ``{Business E-mail Compromise: The 26 Billion Dollar Scam},''
  \url{www.ic3.gov/media/2019/190910.aspx}, September 10, 2019.

\bibitem{Jakobsson2019TheRT}
M.~Jakobsson, ``The rising threat of launchpad attacks,'' \emph{IEEE Security
  \& Privacy}, vol.~17, pp. 68--72, 2019.

\bibitem{phonephish}
A.~Kim, ``{A scam targeting Americans over the phone has resulted in millions
  of dollars lost to hackers. Don't be the next victim},''
  \url{https://www.cnn.com/2019/10/27/business/phishing-bank-scam-trnd/index.html},
  October 27, 2019.

\bibitem{SIADATI201714}
H.~Siadati, T.~Nguyen, P.~Gupta, M.~Jakobsson, and N.~Memon, ``{Mind your
  SMSes: Mitigating social engineering in second factor authentication},''
  \emph{Computers \& Security}, vol.~65, pp. 14 -- 28, 2017.

\bibitem{EmpStudy}
K.~Lee, B.~Kaiser, J.~Mayer, and A.~Narayanan, ``{An Empirical Study of
  Wireless Carrier Authentication for SIM Swaps},''
  \url{https://www.issms2fasecure.com/assets/sim\_swaps-01-10-2020.pdf},
  January 10, 2020.

\bibitem{SoHey}
A.~Greenberg, ``{So Hey You Should Stop Using Texts for Two-Factor
  Authentication},''
  \url{https://www.wired.com/2016/06/hey-stop-using-texts-two-factor-authentication/},
  June 26, 2016.

\bibitem{bypass}
J.~Cox, ``{How Hackers Bypass Gmail 2FA at Scale},''
  \url{https://www.vice.com/en\_us/article/bje3kw/how-hackers-bypass-gmail-two-factor-authentication-2fa-yahoo},
  December 19, 2018.

\end{thebibliography}

\end{document}